\documentclass[journal]{IEEEtran}
\usepackage[section] {placeins}
\usepackage{graphicx}
\usepackage{amsmath}
\usepackage{subfigure}
\usepackage{rotating}
\usepackage{multirow}
\usepackage{array}
\usepackage{rotating}
\usepackage{auto-pst-pdf}
\usepackage{graphicx}
\usepackage{ esint }
\usepackage{amssymb}

\begin{document}

\title{A Survey of Home Energy Management Systems\\ in Future Smart Grid Communications}

\author{I. Khan$^{1}$, N. Javaid$^{1,2}$, M. N. Ullah$^{1}$, A. Mahmood$^{2}$, M. U. Farooq$^{2}$\\\vspace{0.4cm}

$^{1}$Dept of Electrical Engineering, COMSATS Institute of IT, Islamabad, Pakistan.\\
$^{2}$CAST, COMSATS Institute of IT, Islamabad, Pakistan.}

\maketitle

\begin{abstract}
\boldmath
In this paper we present a systematic review of various home energy management (HEM) schemes. Employment of home energy management programs will make the electricity consumption smarter and more efficient. Advantages of HEM  include, increased savings for consumers as well as utilities, reduced peak to average ratio (PAR) and peak demand. Where there are numerous applications of smart grid technologies, home energy management is probably the most important one to be addressed. Utilities across the globe have taken various steps for efficient consumption of electricity. New pricing schemes like, Real Time Pricing (RTP), Time of Use (ToU), Inclining Block Rates (IBR), Critical Peak Pricing (CPP) etc, have been proposed for smart grid. Distributed Energy Resources (DER) (local generation) and/or home appliances coordination along with different tariff schemes lead towards efficient consumption of electricity. This work also discusses a HEM system's general architecture and various challenges in implementation of this architecture in smart grid.

\end{abstract}


\IEEEpeerreviewmaketitle
\section{Introduction}

\IEEEPARstart{E}{lectrical} power grid is a system with some or all of the following four capabilities, power generation, transmission, distribution and control. Integration of advanced Information and Communication Technologies (ICT) increases the efficiency of the traditional grid which makes it capable to make decisions fast and accurate. The integration of ICT in traditional grid results in more automation, reliable provision of electrical services, safe operation of electrical appliances and hence an increased level of consumer comfort~\cite{khan2013home}. The advent of smart grid has argued the proposal of several emerging technologies in past decade from many researchers across the globe. Smart meters, Advanced Metering Infrastructure (AMI), bidirectional communication, home automation and Home Area Networks (HANs) are the technologies addressed by various researchers~\cite{erol2011wireless}. Traditional power grid has been serving humanity for the last 100 years. Population across all over the world and the dependency level of human on electricity are continuously and exponentially increasing phenomena. As not much change has been made to the traditional grid to cope with the increased demand, the ultimate result is that the traditional grid has worn out and the idea of smart grid has evolved~\cite{kailas2012survey}.\\
\indent Distributed applications for smart grid may be found  in  consumption, distribution, transmission, and generation of electrical energy. Smart grid enhances the electricity usage efficiency. If home appliances are equipped with sensors, AMI may be used for load prediction of a specific area~\cite{anas2012minimizing}. Efficient consumption of electricity proves beneficial to us both socially and economically. The employment of HEM systems in a residential area reduces energy bills for consumers and peak demand. With a normal demand in peak hours the utilities are able to provide power from base plants and hence contribution of Green House Gases (GHG) is less towards environmental pollution. In~\cite{di2009cooperative}, a power quality monitoring strategy has been enabled by using sensor networks in smart grid (transmission \& distribution application). In~\cite{javaid2012monitoring}, ZigBee  protocol has been used for monitoring and controlling of power for efficient consumption and distribution (consumption and distribution application). Distributed power generation option is always there in smart grid technology, where in-home electricity can be generated (photovoltaic, wind power), use the required energy locally and sale spare power back to utility.\\
\indent The load demand curve in traditional grid, where flat pricing rates are active, shows that load demand is comparatively high during peak periods when compared to off-peak. So the utilities are not able to provide such high power from their base plants (Hydal power stations) and they have to compulsively switch on their peaker plants (thermal power plants) for which the power generation costs and emission of GHG,s are very high as compared to base plants. The originally inelastic load demand curve needs to be altered to reduce peak load demand, energy cost and emission of GHGs. A HEM system in smart grid enables Demand Response (DR) and Demand Side Management (DSM) programs. DR programs help in managing and altering electricity consumption on electricity supply basis. Whereas DSM programs are related to planning, implementation and evaluation policies and techniques which are formulated to alter the electricity usage of consumers. Different optimization methods, protocols and standards have been proposed for efficient coordination of domestic appliances and DER to reduce peak load and energy usage charges. A continuous work in this regard is underway across the globe, at academic, industrial and at government level.\\
\indent Two sorts of HEM schemes are discussed in this paper i.e. one is communication based and the other is optimization based. The HEM schemes are combined with different pricing schemes in order to make the scheme more efficient. For example in~\cite{baig2013smart}, a day ahead pricing has been used in a HEM scheme to minimize the electricity charges of a consumer.\\

\hfill
\section{Home Energy Management and Monetary Cost Minimization}
Energy management is a term, which has been applied with various meanings in different situations. It is a broader term but our concern is with energy saving in  business, public sector/ government organizations and homes. The process of observing, controlling and conserving electricity usage in an organization/ building is  termed as energy management or home energy management~\cite{kailas2012survey}. It has been reported that 40$\%$ of the global power consumption takes place inside residential buildings~\cite{Ullah2013survey}. In the case of smart grid the consumers are able to generate local energy (in-home energy) from distributive generation units. Also there is an opportunity for various pricing schemes, so the need for HEM programs has been addressed by many researchers. Previous work has proved that energy management programs with feed-in (local generation) give increased savings as compared to without feed-in. Fig. 1~\cite{erol2011wireless}, shows the savings for both feed-in and without feed-in systems.
\begin{figure}[!h]
\centering
\includegraphics[height=6cm, width=8cm]{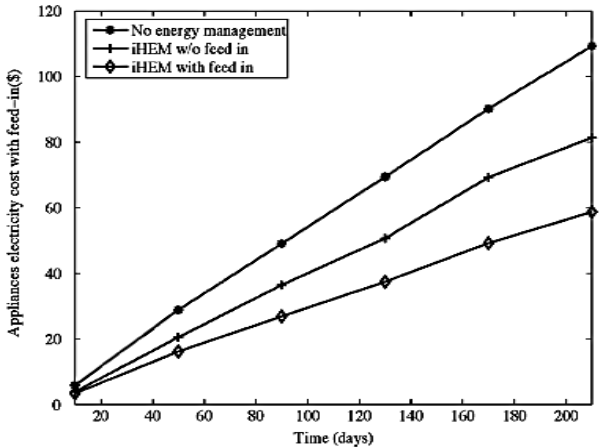}
\caption{Appliance electricity costs with feed in }
\end{figure}\\
\indent Various pricing schemes have been employed for billing purposes by the distribution companies to make an energy management scheme more efficient. The pricing schemes proposed so far for smart grid are RTP, ToU pricing scheme, CPP, day ahead pricing (DAP), etc. In RTP scheme, consumer is informed about the pricing rates at hourly basis as the rates may change hourly. In ToU pricing scheme a consumer is charged least during off-peak, less during mid-peak and more during peak periods.\\
\indent Base plants which are usually categorized as renewable energy sources like hydro power plants provide power to base loads. For the peak periods when the consumer demand climbs rapidly the utilities bring their peaker plants online to maintain balance between load and demand. The peaker plants are run by diesel generators hence the generation costs are comparatively high. Also, emission of high amount of GHG is associated with the operation of these plants~\cite{erol2011wireless}. The overall effect of this process is, increase in electricity prices and global environmental problems. The energy management algorithms can shift load from peak periods to off-peak periods to avoid the services of peaker plants and hence reduce the generation cost and emission of GHG. In this regard, the previous work shows that the objective of avoiding the services of peaker plants can also be achieved by scheduling the DER i.e. scheduling the DER by optimizing an objective function.\\
\indent DSM concept was first introduced in the late 1970s~\cite{costanzo2012system}. A DSM program contributes in reduced emissions of GHG, reliable provision of electricity and reducing the energy cost. Traditional grid has DSM programs for consumers like industrial plants and commercial buildings; however it does not offer any such program for residential consumers. The reasons behind this are  lack of sensors, effective automation tools and efficient communication. Also the advantages for several DR programs are negligibly small when compared with its implementation costs. However in smart grid, the set of smart meters, low cost sensors, smart loads, and the integration of ICT has opened a window for residential energy management programs~\cite{erol2011wireless}. In this regard a lot of work is in progress for designing efficient routing protocols for WSNs to address the issues of efficient energy utilization, delay, path loss, interference and quality of service etc. In~\cite{ahmad2013density}-~\cite{aslam2012ceec} different energy efficient routing protocols have been proposed by the authors.\\
\indent In subsections below few of energy management schemes are presented. The basic aim of these schemes is to reduce peak load demand, electricity consumption charges and the emission of GHG.

\subsection{Optimization-Based Residential Energy Management (OREM)}
In ~\cite{erol2011wireless} a linear programming (LP) model has been proposed by the authors that aim on minimizing the cost of electricity at home. The scheme assumes that a day is divided in equal length, consecutive time slots with different prices of electricity similar to ToU tariff. The proposed objective function makes sure to reduce the home energy expenses by scheduling the home appliances in appropriate time slots. Input for LP model is the consumer requests and the model gives optimum appliance scheduling at the output.\\
\indent The objective function proposed is defined as~\cite{erol2011wireless}\\
\begin{equation}
\sum_{i=1}^{I}\sum_{j=1}^{J}\sum_{t=1}^{T}\sum_{k=1}^{K}E_iD_iU_tS_{t}^{ijk}
\end{equation}
Where\\\\\\I    Number of appliances\\\\J    Days\\\\K Number of requests\\\\T    Number of time slot\\\\$E_i$    Energy consumption of appliance $\dot{\imath}$\\\\$D_i$    Length of cycle of appliance $\dot{\imath}$\\\\$U_t$    Unit price for slot t\\\\${S_t^{ijk}}$    The ratio of the time that an appliance $\dot{\imath}$ runs in a time slot 't' to the total length of appliance cycle.\\
\indent Scheduling an appliance in an appropriate time slot may bring a non acceptable amount of delay to the appliance cycle and an exploded load in the low price time slots. To tackle this problem an upper bound delay $D_{max}$ is specified for each appliance which is less than or equal to the length of two time slots. Mathematically\\
\begin{equation}
D_{max}\leq 2D_i
\end{equation}\\
Where $D_{i}$ denotes the operation cycle of appliance $\dot{\imath}$.\\
\subsection{In-Home Energy Management (iHEM)}
iHEM, an energy management scheme  for domestic energy management is presented in~\cite{erol2011wireless}. The scheme uses smart appliances, a central energy management unit (EMU) and wireless sensor home area networks (WSHANs) for communication purposes among appliances, EMU and smart meters. iHEM uses Zigbee protocol for the implementation of wireless sensor network, organized in cluster-tree topology. The application is based on appliance coordination system (ACS). Unlike OREM, the consumer's demands are processed in near real time in iHEM.\\
\indent A consumer may turn on any appliance at any moment on the clock irrespective of the peak hours concern and iHEM suggest a convenient start time to the consumer. On switching the appliance on, a START-REQ packet is sent by the appliance to the EMU. Upon receiving the START-REQ packet, EMU communicates with the storage system to inquire about the available stored energy by sending AVAIL-REQ packet. EMU also communicates with smart meter to know about the updated prices. The storage unit sends an AVAIL-REP packet in reply, containing the information about the amount of stored energy. When EMU receives the AVAIL-REP packet, it schedules a convenient start time for the appliance according to the iHEM algorithm and notifies it to consumer by sending a START-REP packet. The consumer at this stage may be willing to negotiate with EMU, through the NOTIFICATION packet. Message flow is shown below in Fig. 2~\cite{erol2011wireless}, for iHEM application.
\begin{figure}[!h]
\centering
\includegraphics[height=6cm, width=8cm]{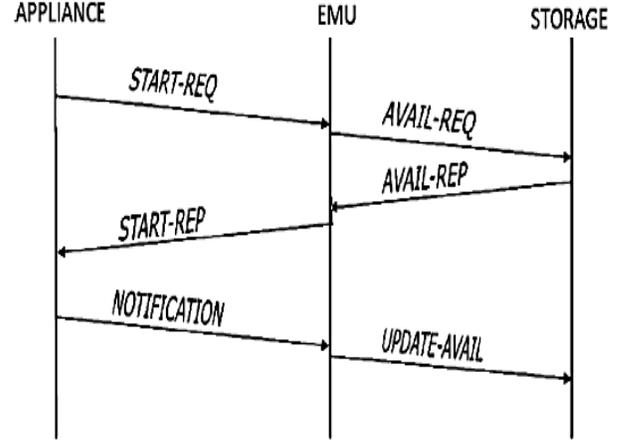}
\caption{Message flow in iHEM application}
\end{figure}
30\% of the load takes place during peak hours in  the absence of energy management programs. By employing iHEM, peak load can be reduced up to 5\%~\cite{erol2011wireless}. iHEM also reduces carbon emission and energy consumption costs.\\
\subsection{Appliance Coordination (ACORD)}
In~\cite{erol2010wireless} ACORD scheme has been proposed to benefit from ToU pricing and decrease energy cost. Aim of ACORD scheme is to shift the consumer load to off-peak periods. In-home WSN’s are used for delivery of consumer requests to EMU. The work shows that the rate of consumer requests has a sizeable effect on energy cost reduction. Energy consumption lowers significantly with an increase in request rates from consumer side. Alternatively, consumer's participation in the energy management program enhances the efficiency of the scheme. This scheme only considers the scheduling of home appliances.\\
\subsection{Optimal and Automatic Residential Energy Consumption Scheduler}
 The optimization based residential load control scheme discussed in~\cite{mohsenian2010optimal}, is based on simple LP computations. The scheme is proposed for real time pricing which needs a price predictor. The combination of price predictor and energy consumption scheduling (ECS) device significantly lowers the PAR in load demand for different load scenarios. The optimization problem~\cite{mohsenian2010optimal}, given below is solved by using LP techniques.\\
\begin{equation}
\begin{split}
\underset{x\in \mathcal{X},v^h \forall h\in \mathcal{H}}{\text{minimize}}\sum_{h=1}^{H}\ v^{h}+\lambda_{wait}\sum_{h=1}^{H} \sum_{a\in \mathcal{A}} (\delta_{a})^{\beta_a-h} x_{a}^{h}/E_{a} \\
a^h\sum_{a\in \mathcal{A}}x_{a}^{h}\leq v^h ,\hspace{0.1cm}  \forall h\in \mathcal{P},\\
b^h\sum_{a\in \mathcal{A}}x_{a}^{h}+(a^h-b^h)c^h\leq v^h ,\hspace{0.1cm}  \forall h\in \mathcal{P},\\
\hat{a}^h\sum_{a\in \mathcal{A}}x_{a}^{h}\leq v^h ,\hspace{0.1cm}  \forall h\in \mathcal{H}/\mathcal{P},\\
\hat{b}^h\sum_{a\in \mathcal{A}}x_{a}^{h}+(\hat{a}^{h} - \hat{b}^{h})\hat{c}^{h}\leq v^h ,\hspace{0.1cm}  \forall h\in \mathcal{H}/\mathcal{P}.\\
\end{split}
\end{equation}\\
where, ${v^h}$ is an auxiliary variable, $\lambda_{wait}$ shows the importance of waiting cost term in objective function, $\delta_{a}$ is the cost of waiting, $\alpha_{a}$ is the scheduling starting time of appliance a, $\beta_{a}$ is the scheduling ending time of appliance a, $x_{a}^{h}$ is the energy consumption scheduling vector, and $E_{a}$ is the predetermined energy of appliance a.

The simulation results shown in Fig. 3~\cite{mohsenian2010optimal}, show the reduction in PAR by solving the optimization problem by LP techniques. The results are satisfactory for utilities.
\begin{figure}[!h]
\centering
\includegraphics[height=6cm, width=8cm]{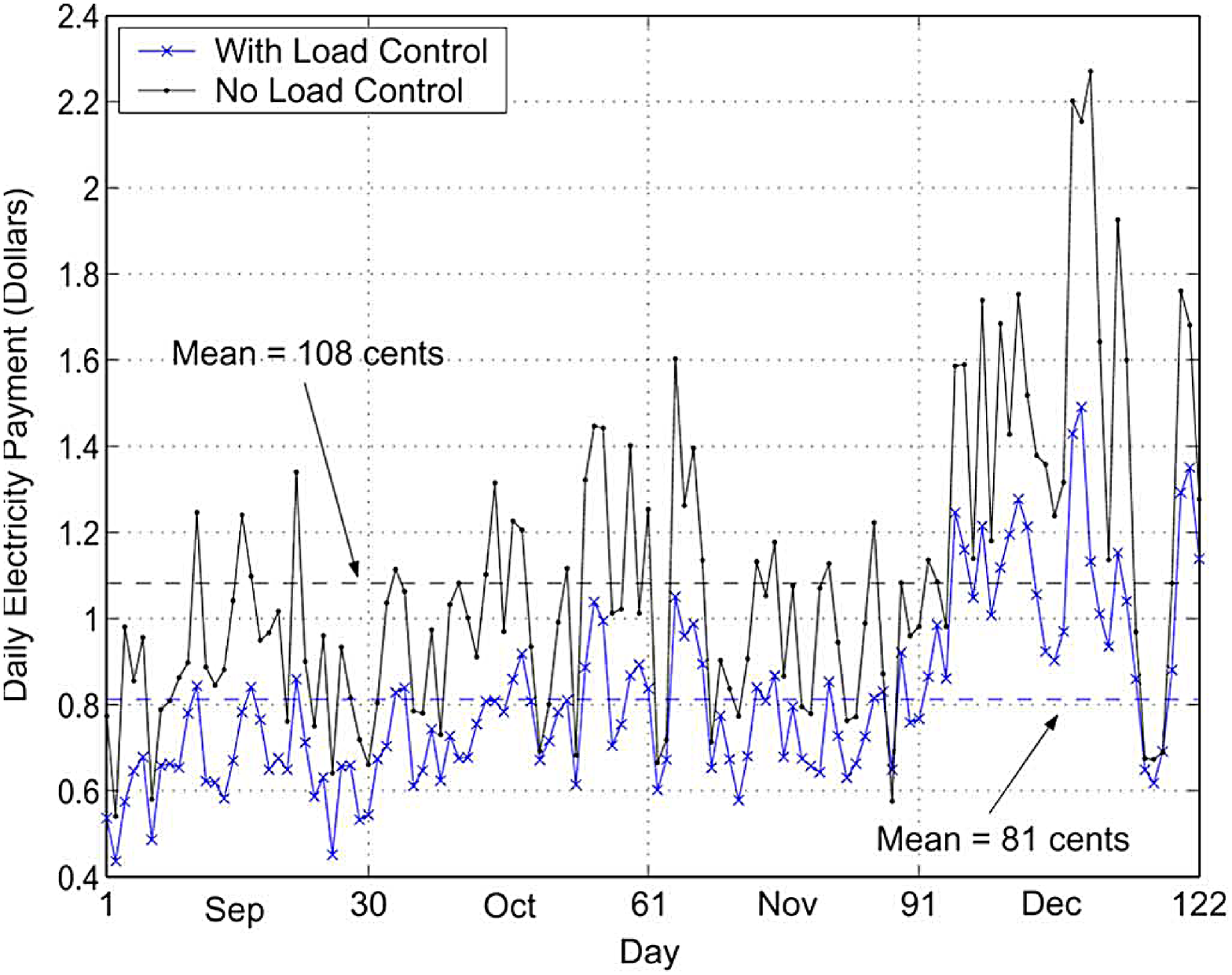}
\caption{Trends of daily PAR for a typical residential load based on DAP adopted by IPC from 1 September to 31 December 2009 }
\end{figure}\\
\subsection{Appliance Coordination with Feed-In (ACORD-FI)}
ACORD-FI~\cite{erol2010using}, is another energy management scheme for energy-aware smart homes. In ACORD-FI both the home appliances and distributed energy resources are scheduled with the purpose of reducing the energy bill and GHG~\cite{erol2010using}. ACORD-FI schedules consumer requests considering peak hours, local energy generated and other conflicting requests. ACORD-FI uses WSNs for communication between EMU and appliances and smart meters.\\
\subsection{Optimum Load Management (OLM) Strategy}
In~\cite{lujano2012optimum}, an optimization based residential load management strategy has been proposed. The optimization problem needs several interests forecasting and activity scheduling by users to form an objective function. Various interests are local power production i.e. from solar, wind etc, load, and electricity prices for next day. Following objective function is produced as a result~\cite{lujano2012optimum}.
\begin{equation}
\begin{split}
\fint=\sum_{i=1}^{i=24}[(\sum_{n=1}^{n=N}VUA_n{(i)}PDCA_n{(\alpha_n{,i})}+\\ \sum_{k=1}^{k=K}VUEV_k(i)PDCEV_k{(\beta_k{,i})})-\\EP(i)(\sum_{n=1}^{n=N}PDCA_n{(\alpha_n{,i})}+\\
\sum_{k=1}^{k=K}PDCEV_k{(\beta_k{,i})})+EP(i)(WP(i)+PVP(i))
\end{split}
\end{equation}\\
Where\\
$\fint$ is the difference between the amount the user have paid and the cost of obtaining the required energy from grid. $VUA_n{(i)}$ is the $\imath th$ hour value of  appliance of user n, $PDCA_n{(\alpha_n{,i})}$ shows the power demand of user n for $\imath th$ hour, $VUEV_k(i)$ represents the user value to travel in electric vehicle k for $\imath th$ hour,$PDCEV_k{(\beta_k{,i})})$ is the power demand of vehicle k for charging purposes in $\imath th$ hour. $PVP(i)$ and $WP(i)$ are forecasted photovoltaic and wind power for $\imath th$ hour respectively. $EP(i)$ is the forecasted electricity consumption prices for $\imath th$ hour.
The authors proposed heuristic optimization techniques to solve the optimization problem due to the nonlinear nature of the function. Although these techniques do not ensure to find global best solution but can give fairly good solutions with low computational time~\cite{lujano2012optimum}.Simulation results show that OLM can reduce energy bill by 8-22\%~\cite{lujano2012optimum}.\\

\subsection{Decision support Tool (DsT)}
Decision Support Tool (DsT) has the primary aim to help  users in making intelligent decisions during their appliances operation. Advantages of energy management program may be increased if DER coordination is adopted in parallel with appliance coordination. In~\cite{pedrasa2010coordinated}, the concept of DER coordination has been evaluated. The work has used an enhanced particle swarm optimization (PSO) solver i.e. CPSO-R, to quantify value added by the DER coordination. Coordination value has been calculated first for the case when each DER is scheduled independently and then for the case when the DER cooperates with each other. A typical smart home case study is shown in the Fig. 4~\cite{pedrasa2010coordinated}.
\begin{figure}[!h]
\centering
\includegraphics[height=6cm, width=8cm]{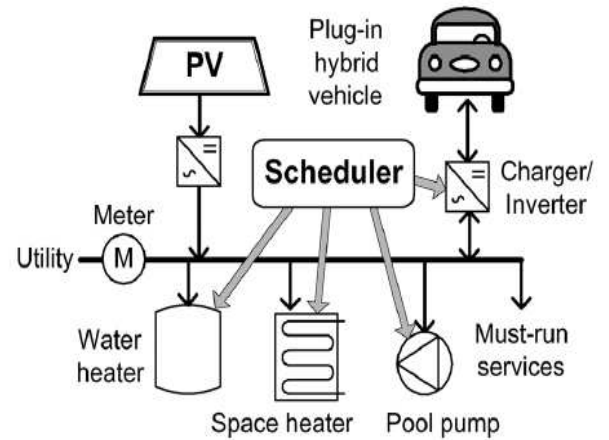}
\caption{A DsT smart home case study}
\end{figure}
DsT is composed of DER scheduling algorithm and an energy service model. The net benefit of the consumer is maximized by scheduling the controllable DER according to the scheduling algorithm. And the consumer energy bill is reduced by 16-25\%~\cite{pedrasa2010coordinated}.

\subsection{Sensors and Control System}
The devices included in this subsystem constitute the basic part of HEM systems. In future smart homes, these devices will provide benefits in the form of facilitating local power generation, managing energy storage centers and diagnostic at a micro-level. Besides power detector capability the sensors are also used for sensing an environmental phenomenon like humidity, temperature or even for inquiring the absence or presence of people in a room. A controller is necessary for receiving the remote control commands to control the operation of a home appliance. Two main issues with deployment of these devices are accuracy and compatibility.\\
\subsection{Monitoring and Control Devices}
Monitoring and control devices provide a visual interface to the users. Data injected from IPMR is visible to the consumers via an in-home display (IHD), where a user can see their real-time consumption and electricity prices at that time. The two challenges for monitoring and controlling devices are user friendliness and simple integration of control interface. Usually incorporation of too many appliance controls on a single control panel may result in a panel with large number of control buttons which will never be a feasible design especially for kids and senior citizens. Similarly the integration of appliances and devices from different vendors and hence using different standards must be an effective and an open research issue~\cite{kailas2012survey}.\\
\subsection{Intelligent Power Management Rostrum (IPMR)}
IPMR is considered to be the heart of HEM system. The primary purpose of integration of IPMR is the exploitation of data from sensors, external internet (utility company) and local environment and transfer it to the IHD for user's assistance. Alternatively IPMR provides home automation. To be more specific IPMR provides three kinds of services, power management services, context aware services and social network services.\\
\section{Challenges for Smart Grid}
Smart grid is smart because of the fast communication and efficient networking capabilities. A lot of work has been done in this regard to make smart grid to get an entry on the stage and serve humanity everlastingly. Still there are some challenges to be tackled before smart grid can do its job. We present some major challenges related to smart grid in this part of the work. These challenges will channelize the research directions for future smart grid.\\
\subsubsection{Interoperability}
When equipments, devices or appliances having different communication and networking technologies can communicate effectively, interoperability is satisfied. Different communication technologies may be adopted by different utility companies, vendors and users~\cite{kailas2012survey}. Therefore it becomes essential to satisfy interoperability so that a number of heterogeneous communication and networking technologies could coexist in various parts of smart grid. For example an Energy Management system may use WiFi and ZigBee for communication purposes. A lot of work can be done in this context.\\
\subsubsection{Scalability}
A system whose performance increases with addition of more hardware proportionally to the number of hardware added is said to be a scalable system. Smart grid like traditional grid will involve a large number of users and the number will increase every moment, hence scalability becomes an issue. Tests for scalability made on small scale may not be valid when used with such a huge amount of user. One approach towards scalability problem is using sensor networks which give comparatively good scalability results.\\
\subsubsection{Interdisciplinary}
Smart grid involves different stakeholder (societies,organizations and systems). Hence the research area has been an interdisciplinary in nature which has always been a tough job. In smart grid, one can see the integration of power systems with actuation, security, control, communication and networking system.\\
\subsubsection{Security and Privacy}
Where there is interconnection of two systems or networks (wired or wireless), there are issues of security and privacy and the same is true in the case of smart grid. Threat of cyber vulnerabilities for future smart grid is an important issue to be tackled. The security issues in smart grid may include accessing smart meter data in an unauthorized fashion, electricity theft, accessing of home appliances control by an unauthorized person or system, or attacking smart grid to affect power continuity. An outsider can access smart metering data and sensible information may be leaked out. This issue is a major hinder in modernizing the traditional grid~\cite{kailas2012survey}.

\section{Conclusion}
On a conclusion note, in this work we have revisited the need for home energy management for efficient usage of electricity in smart grid. Efficient consumption of electrical energy results fruitfully in lowering peak load, reducing electricity bills and minimizing the emission of GHG. Smart grid is facilitated by bidirectional communication and effective home automation hence an intelligent home energy management system can be designed. Our work has discussed several HEM schemes,summarized in table. 1, where different pricing schemes have been applied to get social and economical advantages. Both optimization-based and communication-based HEM techniques have been

\begin{table*}[t]
\centering
\caption{\bf Comparison of Different Home Energy Management Schemes}
\begin{tabular}{|m{1.5cm}|m{1cm}|m{1.7cm}|m{1.5cm}|m{1.6cm}|m{1.2cm}|m{1.3cm}|m{1.3cm}|}
\hline
\textbf{Scheme Name} & Pricing & Goal &  Method & Communication & Coverage & Peak Load Reduction & Monthly Bill Reduction \\
\hline
\textbf{OREM} & {ToU} & Cost minimization & LP Based Optimization & No & Local & N/A & 35\% \\
\hline
\textbf{iHEM} & {ToU} & Cost minimization & Interactive Load Shifting & Yes & Local & 40\% & 30\% \\
\hline
\textbf{RLC} & {RTP} & Cost and PAR minimization & LP Based Optimization & No & Local & 22\% & 16-25\% \\
\hline
\textbf{OLM} & {RTP} & Cost and Consumption minimization & Heuristic Optimization Techniques & No & Local  & N/A & 8-22\% \\
\hline
\textbf{DsT} & {ToU,CPP} & Scheduling DER & PSO & No & Local & N/A & 16-25\% \\
\hline

\end{tabular}
\label{tab:template}
\end{table*}

evaluated. We have also discussed a general HEM system's architecture and the challenges that future smart grid will be facing. We are of the hope that this work will channelize the efforts towards a more efficient, user friendly HEM system for future smart grid.

\begin{figure}[!h]
\centering
\includegraphics[height=6cm, width=8cm]{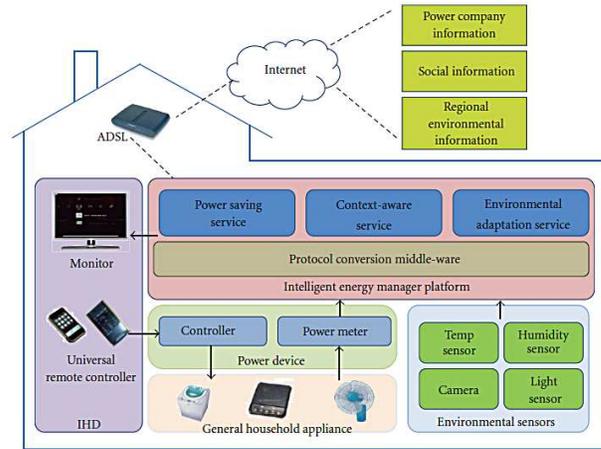}
\caption{HEM System Architecture for Future Smart Grid}
\end{figure}

\end{document}